\documentclass[aip,cha,reprint,twocolumn]{revtex4-1}

\pdfoutput=1

\usepackage{graphicx}%
\usepackage{subfigure}%
\usepackage{amssymb}%

\newcommand{\kb}{k_{\mathrm{B}}}
\newcommand{\rhom}{\rho_{\mathrm{m}}}

\newcommand{\rhoc}{\rho_{\mathrm{c}}}
\newcommand{\ud}{\mathrm{d}}
\newcommand{\ple}{P^{(\mathrm{leq})}}
\newcommand{\peq}{P^{(\mathrm{leq})}}
\newcommand{\nub}{\nu^{(\mathrm{B})}}
\newcommand{\kappab}{\kappa^{(\mathrm{B})}}
\newcommand{\nuj}{\nu^{(\mathrm{J})}}

\newcommand{\nuw}{\nu^{(\mathrm{W})}}

\begin{document}
\title{\bf A two-stage approach to relaxation in billiard systems of 
locally confined hard spheres}
\author{Pierre Gaspard}
\email{gaspard@ulb.ac.be}
\author{Thomas Gilbert}
\email{thomas.gilbert@ulb.ac.be}
\affiliation{Center for Nonlinear Phenomena and Complex Systems,\\
Universit\'e Libre de Bruxelles, Code Postal 231, Campus Plaine,
B-1050 Brussels, Belgium}
\begin{abstract}
  We consider the three-dimensional dynamics of systems of many
  interacting hard spheres, each individually confined to a dispersive
  environment, and show that the macroscopic limit of such systems is 
  characterized by a coefficient of heat conduction whose value reduces to 
  a dimensional formula in the limit of vanishingly small rate of interaction. 
  It is argued that this limit arises from an effective loss of memory. 
  Similarities with the diffusion of a tagged particle in binary mixtures are 
  emphasized.
\end{abstract}
\maketitle

\begin{quotation} 
  The derivation of the macroscopic transport equations of hydrodynamics 
  and the computation of the associated coefficients for systems described at 
  the microscopic level by Hamilton's equations of classical mechanics is a 
  central problem of non-equilibrium statistical physics. The periodic Lorentz 
  gas provides an example where this program can be achieved and
  Fick's law of diffusion established \cite{Bunimovich:1980p2706,  
  Bunimovich:1981p479, Bunimovich:1991p47}. Furthermore, by tweaking the 
  system's geometry so that tracer particles hop from cell to cell at nearly 
  vanishing rates, memory effects disappear and the dynamics of tracers
  is well approximated by a continuous time random walk 
  \cite{Zwanzig:1983p33}. In this regime, the diffusion coefficient
  takes on a simple limiting value, given by a dimensional formula
  \cite{Machta:1983p182}, by which we mean that its expression reduces
  to the square of the length scale of the cell separation multiplied
  by the hopping rate. Similarly, it was found that heat transport
  in systems of confined hard disks with rare interactions reduces to
  a stochastic process of energy exchanges which obeys Fourier's law
  of heat conduction, with the coefficient of heat conductivity also
  given by a dimensional formula \cite{Gaspard:2008p332,
    Gaspard:2008p334, Gaspard:2008p341}, where the timescale is that
  of energy exchanges among particles in neighboring cells. Here we
  extend these findings to mechanical systems of confined hard
  spheres, whose stochastic limit was already studied elsewhere
  \cite{Gaspard:2009p988}, and review in some details the analogy with
  the problem of mass transport in the periodic Lorentz gas.
\end{quotation}

\section{Introduction}

A long-standing problem in non-equilibrium statistical mechanics has
been to provide a derivation from first principles of Fourier's law of
heat conduction in insulating materials in the framework of Hamiltonian
mechanics \cite{Bonetto:2000p13477}. A centerpiece of the puzzle is to
identify the conditions under which scale separation occurs so the
time evolution at the microscopic level can somehow be reduced to a
hydrodynamic equation at the macroscopic level, which is an
especially challenging problem for interacting particle systems.

As emphasized in earlier papers \cite{Gaspard:2008p332,
  Gaspard:2008p334, Gaspard:2008p341}, the following two-step
program, which consists of (i) identifying an intermediate level of
description---a mesoscopic scale---where the Newtonian dynamics can be
consistently approximated by a set of stochastic equations, and (ii)
subsequently analyzing the statistical properties of this stochastic
system and computing its transport properties in the hydrodynamic
scaling limit, can be successfully achieved in a class of chaotic
billiard systems composed of many hard disks trapped in a
semi-porous material which prevents mass transfer and yet allows
energy transfer through elastic collisions among neighboring disks.

Under the assumption that collisions among moving disks are rare
compared to wall collision events, the global multi-particle
probability distribution of the system typically reaches local
equilibrium at the kinetic energy of each individual
particle before energy exchanges proceed. This mechanism naturally
yields a stochastic description for the process of energy exchanges in
the system. In a subsequent paper \cite{Gaspard:2009p988}, it was
shown that the same reduction applies to systems of confined hard
spheres. 

Our purpose in this paper is twofold. Our main objective is to
establish a comparison between the process of energy transfer in
models of interacting hard spheres with local confinement rules, on
the one hand, and the diffusive motion of tracer particles in 
low-dimensional billiard tables such as the finite-horizon periodic
Lorentz gas, on the other. Specifically, we show that the two systems,
under equivalent assumptions of local equilibration, whose accuracy
can be precisely controlled by tuning the systems' parameters down to
a critical geometry, are both amenable to stochastic descriptions in
the form of master equations, whether for the distribution of energies
in the case of the former systems, or that of mass in the
latter. Taking the hydrodynamic limit of these stochastic systems,
we obtain explicit values of the transport coefficients of heat 
conduction and diffusion respectively, which turn out to share the
remarkable property that they are given by simple dimensional
formulae, i.~e. by the square of the mesoscopic length scale
multiplied respectively by the rates of energy or mass transfer. 

Turning back to the billiard dynamics, we address our second
objective, which is to show that, under the assumption of separation
of two characteristic timescales, one associated with the local 
dynamics, the other with energy transfers, the process of heat
transport in three-dimensional billiard systems of confined hard
spheres is well-approximated by the corresponding stochastic process
of energy exchanges. We do so by considering the Helfand moment of
thermal conductivity \cite{Helfand:1960p18123} and compute the linear
divergence in time of its mean squared change. Plotting our results as
functions of the system's sizes for different parameter values, we
compute the infinite-size extrapolations to obtain the heat
conductivity and observe a very good convergence to the value
obtained for the stochastic systems for parameter values where an
effective separation of timescales is observed.

The paper is organized as follows. The problem of mass transport in a
spatially periodic billiard table is considered in
Sec.~\ref{sec.mass}. Starting from the pseudo-Liouville equation for
the billiard dynamics, we derive a continuous-time random walk which
describes the stochastic jumps of particles across a lattice, and
analyze its transport properties, comparing it to that of the
billiard. The same procedure is applied in Sec.~\ref{sec.engy2d} to
two-dimensional billiards of confined hard disks. In
Sec.~\ref{sec.engy3d}, we turn to a three-dimensional billiard system
and discuss our numerical results. Conclusions are drawn in
Sec.~\ref{sec.conc}. 

\section{A continuous-time random walk approach to mass
  transport \label {sec.mass}}

\begin{figure}
  \centering
  \includegraphics[width=.45\textwidth]{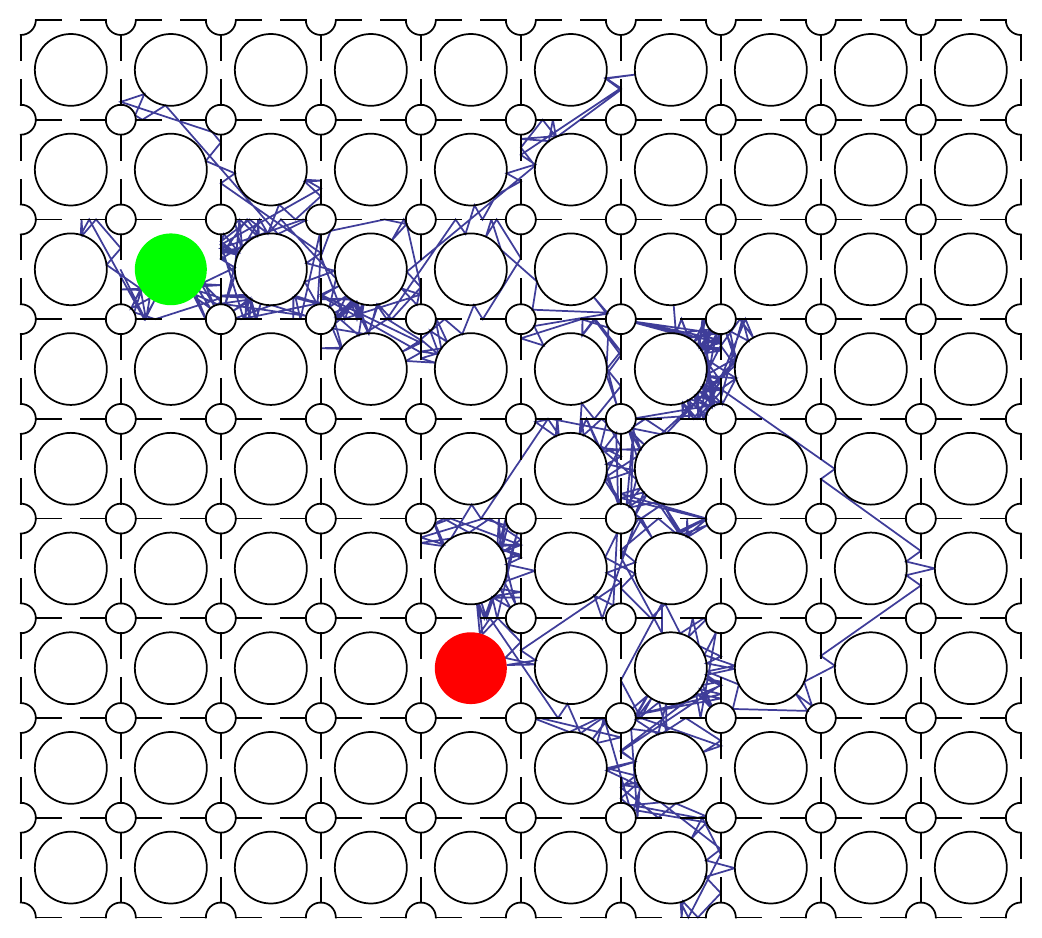}
  \caption{Periodic billiard table on a square lattice with a typical
    trajectory.  Here the central disks in the initial and final cells
    are color-filled. One considers the process of mass transport
    across the periodic cells. The model has several parameters in
    terms of which a dimensional formula of the diffusion
    coefficient is computed.} 
  \label{fig.sinaitable}
\end{figure}

We consider for the sake of the example a periodic array of two-dimensional 
semi-dispersing Sinai billiard tables in the form of square billiard
cells bounded by flat walls of sizes $l$, and connected to each other
by small openings of relative widths $\delta$. All the cells are
identical and contain circular obstacles arranged so that there are no
periodic trajectory that avoid these obstacles. Independent pointwise
tracer particles move across this array, performing elastic collisions
on the obstacles and walls. An example is shown in figure
\ref{fig.sinaitable}. The diffusive properties of this model were
studied in some details in Ref.~[\onlinecite{Gilbert:2009p3207}]. There
the attention was on memory effects and their impact on the value of
the diffusion coefficient. Here we will focus on the derivation of the
kinetic prediction of this coefficient, which yields the
aforementioned dimensional formula, disregarding the memory
effects. The conclusions are similar to those obtained by Machta and
Zwanzig \cite{Machta:1983p182} in the framework of the periodic
Lorentz gas. The continuous-time approach we present below is somewhat
similar to that of Zwanzig \cite{Zwanzig:1983p33}.  

The phase-space configuration of this system refers to a single tracer
particle and is specified by the triplet $\{\mathbf{n},
\mathbf{r},\mathbf{v}\}$, where $\mathbf{n} = (n_x, n_y) \in
\mathbb{Z}^2$ denotes the lattice index of the cell where the tracer
is located, $\mathbf{r}$, its position within the cell and
$\mathbf{v}$ its velocity.

Let $p(\mathbf{n}, \mathbf{r}, \mathbf{v}, t)$ denote the probability
distribution of the tracer. Its time evolution is determined by the
pseudo-Liouville operator \cite{Dorfman:1989p173}, which comprises
four different types of contributions, corresponding to as many
different types of events: 
\begin{enumerate}
\item
  Free advection of the tracer inside the cell is accounted for by the term 
  $-\mathbf{v} \cdot \partial_{\mathbf{r}}$;
\item
  Collisions of the tracer with any of the circular obstacles inside the
  cell are determined by the operators $K^{(d)}$,
  where $d$ refers to a specific disk of radius $\rho_d$ at position 
  $\mathbf{q}^d$ :
  \begin{eqnarray}
    \lefteqn{K^{(d)} p (\mathbf{n}, \mathbf{r}, \mathbf{v}, t) 
      =
      \rho_d \int_{\hat{\mathbf{e}}\cdot\mathbf{v}>0}
      \ud\hat{\mathbf{e}}(\hat{\mathbf{e}}\cdot\mathbf{v})}      
    \nonumber\\
    &\times
    \Big[
    \delta(\mathbf{r} - \mathbf{q}^{d} - \rho\hat{\mathbf{e}})
    p(\mathbf{n}, \mathbf{r}, \mathbf{v} - 2
    \hat{\mathbf{e}}(\hat{\mathbf{e}}\cdot\mathbf{v}), t) 
    \nonumber\\
    &- \delta(\mathbf{r} - \mathbf{q}^{d} + \rho\hat{\mathbf{e}})
    p(\mathbf{n},\mathbf{r}, \mathbf{v}, t)
    \Big]\,;
    \label{diskcoll}   
  \end{eqnarray}
\item
  Collisions of the tracer with any of the flat walls of cell $\mathbf{n}$,
  with operator $W^{(j)}$, where $j$ takes on the values
  $j=1,\dots,4$, corresponding to the right, bottom, left and top
  walls: letting $r_x$ and $v_x$ denote the position and
  velocity components of the tracer along the horizontal axis, wall \#1
  is the right wall at position $l/2$ so that, if the wall is continuous
  with respect to the vertical axis, i.~e. $\delta \equiv 0$, the
  collision operator acts according to
  \begin{eqnarray}
    \lefteqn{W^{(1)} p (\mathbf{n}, \mathbf{r}, \mathbf{v}, t)
      = |v_x| \delta(r_x - l/2)}
    \label{wallcoll1}\\	
    &\times      
    \Big[\theta(-v_x)
    p (\mathbf{n}, \mathbf{r}, -v_x, v_y, t) - \theta(v_x)
    p (\mathbf{n}, \mathbf{r}, \mathbf{v}, t)\Big] \,,
    \nonumber
  \end{eqnarray}
  with similar expressions for $W^{(2)}$, $W^{(3)}$, and $W^{(4)}$;
\item Jump events between neighboring cells take place when a wall
  collision event occurs at a position where the wall is actually
  open, which amounts to subtracting the action of the wall operator
  (\ref{wallcoll1}) above, keeping track of the cell indices: using the
  same set of four indices as above for every wall and assuming that wall
  \#1 has a slit of width $\delta$ centered about $y = 0$, the jump
  operator $J_\delta^{(1)}$ is
  \begin{eqnarray}
    J_\delta^{(1)} p (\mathbf{n}, \mathbf{r}, \mathbf{v}, t)
    = |v_x| \delta(r_x - l/2) \Theta_\delta(r_y) \theta(- v_x)
    \label{jumpcoll1}\\
    \times \Big[
    p (\mathbf{n} + (1,0), \mathbf{r}, \mathbf{v}, t)
    - p (\mathbf{n}, \mathbf{r}, - v_x, v_y, t) 
    \Big] \,,
    \nonumber
  \end{eqnarray}
  where $\Theta_\delta(x) = 1$ if $|x| \le \delta/2$ and $0$
  otherwise.
\end{enumerate}
Collecting these terms together, we obtain the time evolution of 
$p(\mathbf{n}, \mathbf{r}, \mathbf{v}, t)$, given by the
pseudo-Liouville equation: 
\begin{eqnarray}
  \lefteqn{
    \partial_t p (\mathbf{n}, \mathbf{r}, \mathbf{v}, t) =}
  \label{pseudoliouville}	\\
  & \left\{-\mathbf{v} \cdot \partial_{\mathbf{r}}
    + \sum_{d} K^{(d)} + \sum_{j} W^{(j)} 
  \right\} p (\mathbf{n}, \mathbf{r}, \mathbf{v}, t)
  \nonumber\\
  &+ \sum_{j} J_\delta^{(j)} p (\mathbf{n}, \mathbf{r}, \mathbf{v}, t)\,.
  \nonumber
\end{eqnarray}
The terms on the right-hand side of this equation are grouped so as to
distinguish the \emph{local} terms, which do not act on the cell index
in the distribution, from the \emph{non-local} ones, i.~e. the jump
terms, which act on the cell index. Also note that the velocity
amplitude is invariant under every term of this equation and is
therefore a conserved quantity which plays no role once it has been
fixed by the initial condition. 
 
It it an immediate consequence of the ergodicity of the local
dynamics that the following
distribution,
\begin{equation}
  \peq(\mathbf{n}, t) = \int \ud\mathbf{r} \int \ud \mathbf{v} 
  \delta(v^2 - v_0^2) p (\mathbf{n}, \mathbf{r},\mathbf{v}, t)\,,
  \label{loceq}
\end{equation}
is invariant under the local terms of the pseudo-Liouville equation
(\ref{pseudoliouville}). We thus refer to $\peq(\mathbf{n}, t)$ as the
\emph{local equilibrium distribution}. Its time evolution occurs
through jump events only~: 
\begin{eqnarray}
  \lefteqn{\partial_t \peq (\mathbf{n},t) =}
  \nonumber\\
  &  \quad\sum_j
  \int \ud\mathbf{r} \int \ud \mathbf{v} \delta(v^2 - v_0^2)
  J_\delta^{(j)} p (\mathbf{n}, \mathbf{r}, \mathbf{v}, t).
  \label{peqevol1}
\end{eqnarray}
In order to close this equation for $\peq(\mathbf{n}, t)$, we make the
\emph{local equilibrium approximation},
\begin{equation}
  p(\mathbf{n}, \mathbf{r}, \mathbf{v}, t) = 
  \frac{1}{\pi A} \delta(v^2-v_0^2)\peq(\mathbf{n}, t) \, ,
  \label{closure}
\end{equation}
where $A$ is the area of the billiard, e.~g. $A = l^2 - \pi (\rho_1^2 +
\rho_2^2)$ in the case of figure \ref{fig.sinaitable} (where $\rho_1$
denotes the radius of the disk at the center of the cell and $\rho_2$
that of the disks at the cell corners).

Plugging Eq.~(\ref{closure}) into (\ref{peqevol1}), the
pseudo-Liouville equation  (\ref{pseudoliouville}) reduces to the
following \emph{continuous-time random walk} 
\begin{equation}
  \partial_t \peq (\mathbf{n},t) = \sum_j
  \frac{v_0 \delta}{\pi A} [\peq(\mathbf{n} + \mathbf{e}_j, t)
  - \peq(\mathbf{n}, t)]\,,
  \label{meq}
\end{equation}
where $\mathbf{e}_j \in \{(\pm 1, 0), (0, \pm 1)\}$.

In the continuum scaling limit, $\vec{r} = l \mathbf{n}$, $l\to 0$,
$v_0 \sim l^{-2}$,  this reduces to a diffusion equation,
\begin{equation}
  \partial_t P(\vec{r}, t) = \mathcal{D} \nabla^2 P(\vec{r},t)\,,
  \label{FPeq}
\end{equation}
with diffusion coefficient
\begin{equation}
  \mathcal{D} = l^2 \nuj,
  \label{diffcoeff}
\end{equation}
here written in terms of the frequency $\nuj$ of a jump event in a 
specific direction, identical for all four directions,
\begin{equation}
  \nuj = \frac{v_0 \delta}{\pi A}.
  \label{defjumprate}
\end{equation}

\begin{figure}[tbhp]
  \centering
  \includegraphics[width=.45\textwidth]{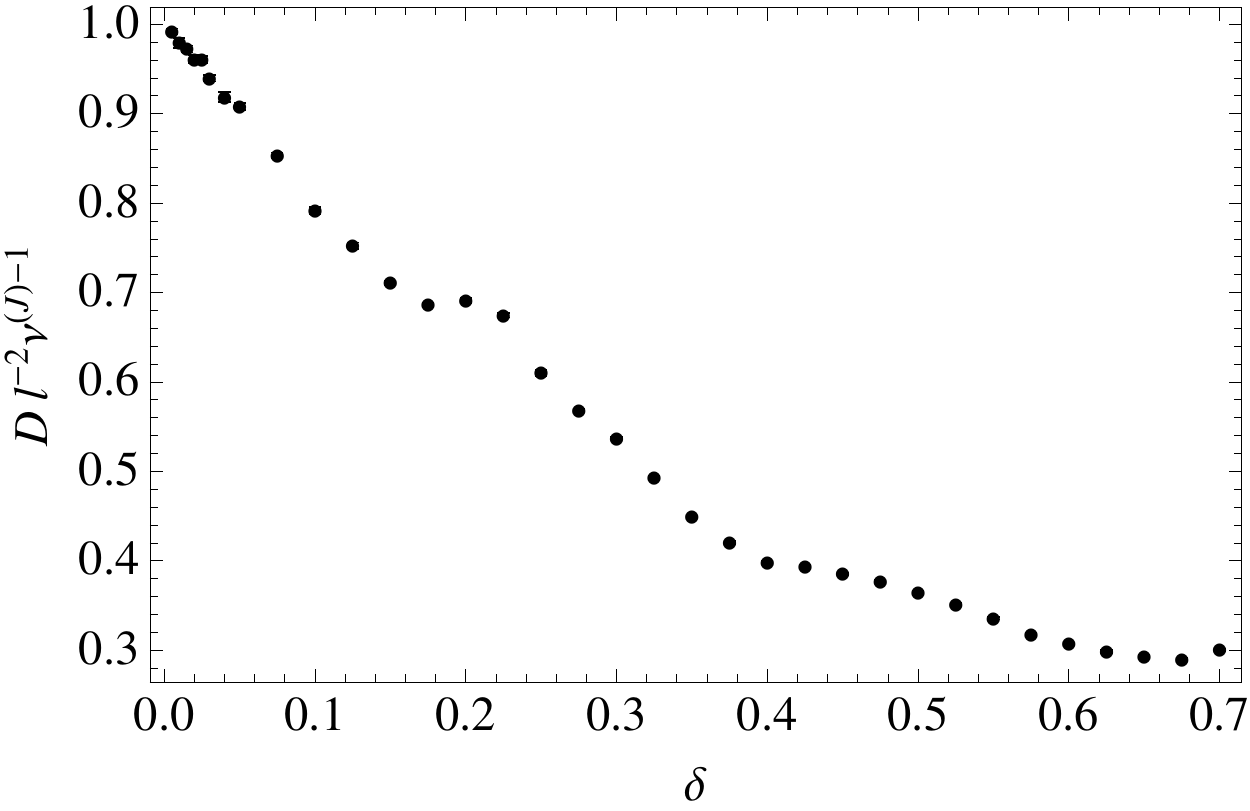}
  \includegraphics[width=.45\textwidth]{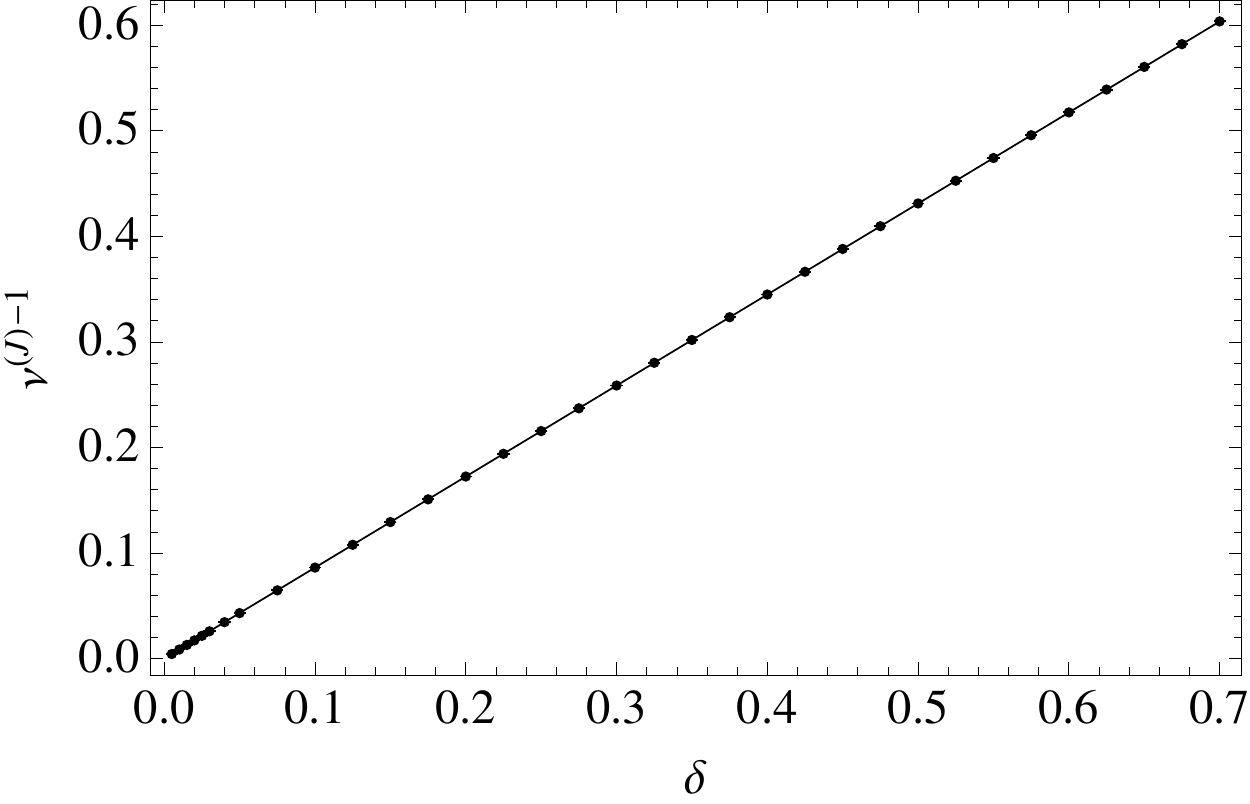}
  \caption{(Top) Diffusion coefficient as measured from the mean-squared
    displacement divided by the number of jumps for the Sinai billiard
    table shown in figure \ref{fig.sinaitable}. (Bottom) Corresponding
    decay time. 
    The dots show numerical measurements and the red line is the formula
    (\ref{defjumprate}).}
  \label{fig.diff}
\end{figure}

Equation (\ref{diffcoeff}) is what we refer to as a \emph{dimensional
formula}: the diffusion coefficient associated with the continuous-time
random walk process (\ref{meq})  is the product between the scale of
displacements squared and hopping rate. The fact that it is an exact
property of the stochastic process is remarkable.

As far as the transport properties of the billiard table go,
Eq.~(\ref{diffcoeff}) is a mere approximation: its validity relies on
the local equilibrium approximation (\ref{closure}). In contrast,
Eq.~(\ref{defjumprate}) is an exact result for the billiard dynamics
as well---valid for all parameter values---as it relies solely on the
ergodicity of the billiard table \cite{Chernov:2001p143}.

The closure approximation (\ref{closure}) is expected to be exact in
the limit of vanishing window widths, $\delta\to0$, whereby the
hopping rate (\ref {defjumprate}) diverges and the diffusion
coefficient (\ref{diffcoeff}) vanishes. In practice however, we may
expect $p(\mathbf{n}, \mathbf{r}, \mathbf{v}, t)$ to converge to the
local equilibrium distribution $\peq(\mathbf{n}, t)$ between
successive jumps so long as the typical 
number of disk collision events within a cell is large. This can be
verified numerically, as shown on the top panel of figure
\ref{fig.diff}. These numerical results were already reported in
Ref.~[\onlinecite{Gilbert:2009p3207}]. Similar results are equally
obtainable for three dimensional billiard systems such as the
three-dimensional periodic Lorentz gas \cite{Gilbert:2011p10752}. 

\section{Heat transport in two-dimensional confining
  billiards \label{sec.engy2d}} 

Similar considerations apply to models of heat transport with mass
confinement \cite{Gaspard:2008p332}. Classes of such models were
initially introduced by Bunimovich \emph{et. al} in
Ref.~[\onlinecite{Bunimovich:1992p621}]. There, the authors proved the
ergodicity of two-dimensional billiard tables consisting of an
arbitrary large number of unit cells placed side by side, each
containing a single disk trapped within the cell's boundaries, but in
such a way that collisions may still take place between disks
belonging to neighboring cells.  

\begin{figure}[thbp] 
   \centering
   \includegraphics[width=.45\textwidth]{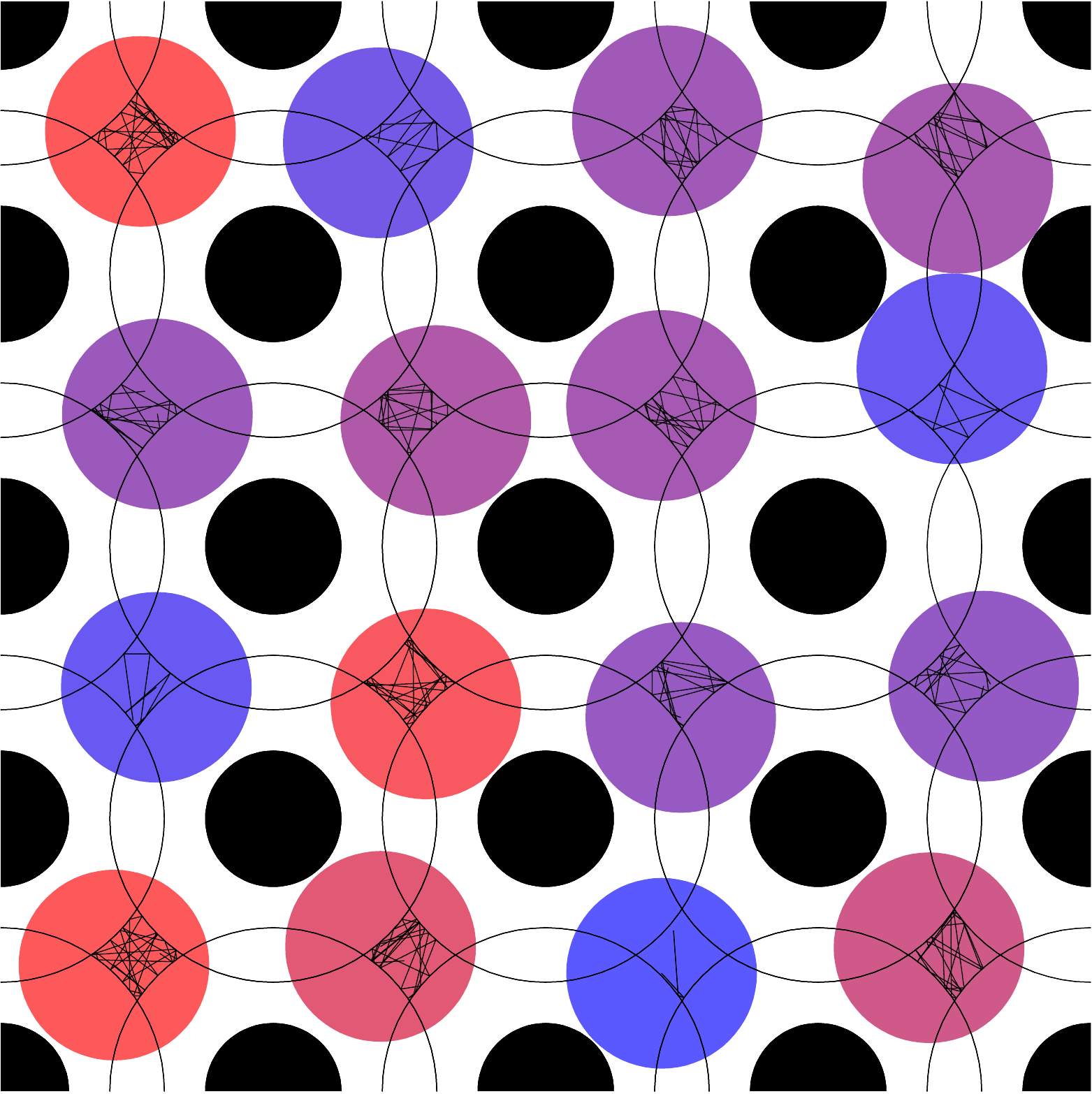} 
   \caption{Example of a two-dimensional billiard table consisting of
     many moving hard disks (colored), each trapped within confining
     walls (black disks), here arranged in a square tiling. The
     centers of the moving disks are bound to the areas delimited by
     the exterior intersections of the black circles around the fixed
     disks (the solid broken lines sample their trajectories). The
     parameters (radii of the black circles $\rho$ and moving disks
     $\rhom$) are so chosen that (i) black circles overlap, so the
     moving disks are confined, and (ii) collisions are allowed (such
     as in the upper right corner), whereby energy exchanges take
     place. The mobile disks are color coded from blue to red with
     growing kinetic energies.}
   \label{fig.2dlatticeb}
\end{figure}

Figure \ref{fig.2dlatticeb} shows an example of such a system
\footnote{For technical reasons pertaining to the probability of
  direction re-collisions between two particles, this example actually
  falls out of the class of systems considered in
  Ref.~[\onlinecite{Bunimovich:1992p621}]. This is unimportant for our
  own considerations.}. In Ref.~[\onlinecite{Gaspard:2008p334}], we
introduce the following parametrization of the model in terms of two
parameters: $\rho$, which characterizes the timescale of wall
collision events, and $\rhom$, which characterizes that of binary
collisions. 

On the one hand, the dynamics within an isolated cell of width $l$
boils down to the motion of a point particle in an area bounded by the
exterior intersection of four disks of radius $\rho$, $l/2 \leq \rho <
l/\sqrt2$. In the absence of interaction with neighboring particles,
the mean free path of a particle is given by $\ell = \pi
|\mathcal{B}_\rho|/| \partial \mathcal{B}_\rho|$, where
$\mathcal{B}_\rho$ and $\partial \mathcal{B}_\rho$ denote respectively
the area and perimeter of the billiard cell. The corresponding wall
collision timescale is obtained by multiplying the mean free path by
the speed of the particle.  

Interactions among neighboring particles  on the other hand take place
provided the radius of the moving particles $\rhom$ is larger than a
critical value,  $\rhom > \rhoc = \sqrt{\rho^2 - l^2/4}$. The
corresponding timescale can be computed and shown to diverge with
$(\rhom - \rhoc)^{-3}$. Letting $\nuw(T)$ and $\nub(T)$ respectively
denote the frequencies of wall collision and binary collision events
measured at equilibrium temperature $T$, the following separation of
timescales is assumed, 
\begin{equation}
  \nub(T)\ll\nuw(T),
  \label{septimes}
\end{equation}
which occurs when $\rhom\to\rhoc$.

Under this assumption, we
showed\cite{Gaspard:2008p332,Gaspard:2008p341} that the heat
conductivity of spatially extended  billiard systems, which are
infinite size limits of systems such as depicted in
Fig.~\ref{fig.2dlatticeb}, reduces, up to a dimensional factor $l^2$,
to the frequency of binary collisions, 
\begin{equation}	
  \kappab(T) = l^2 \nub(T).	
  \label{kappanu}
\end{equation}
Furthermore the heat conductivity scales with the thermal speed,
$\kappab(T) \sim \sqrt T$. 

Although Eqs.~(\ref{diffcoeff}) and (\ref{kappanu}) are very much
alike, the derivation of the latter is much more involved than that of
the former.

Proceeding as in Sec.~\ref{sec.mass}, the first part of the program,
which is to  reduce the pseudo-Liouville equation describing the
time evolution of the billiard system to a continuous-time stochastic
process of energy exchanges, follows closely along lines which led
from Eqs.~(\ref{pseudoliouville}) to (\ref{meq}). Indeed, considering
the time evolution acting on the $N$-particle phase-space distribution
$p_N(\{\mathbf{r}_i,\mathbf{v}_i\},t)$, one readily notices that the
local equilibrium distribution 
\begin{eqnarray}
  \lefteqn{\ple_N(\epsilon_1, \dots, \epsilon_N, t) \equiv}
  \label{loceqdis}\\
  &&
  \quad
  \int \prod_{a=1}^N\ud \mathbf{r}_a\ud \mathbf{v}_a \delta(\epsilon_a
  -m v_a^2/2) p_N(\{\mathbf{r}_i,\mathbf{v}_i\},t)
  \nonumber
\end{eqnarray}
is left unchanged by the advection and wall collision
terms. Therefore, and provided binary collision events are rare with
respect to wall collision events, we can make a closure approximation
similar to Eq.~(\ref{closure}) to obtain the time evolution of the
local equilibrium distribution in the form of a stochastic process. 

The result is a master equation which describes the time evolution of
the $N$ cell system with energy variables $\{\epsilon_1, \dots,
\epsilon_N\}$ in terms of energy exchanges between neighboring cells
at respective energies $\epsilon_a$ and $\epsilon_b$ of amount $\eta$,
specified by a stochastic kernel $W$: 
\begin{widetext}
  \begin{eqnarray}
    \partial_t \ple_N(\epsilon_1, \dots, \epsilon_N, t) &=&
    \frac{1}{2}\sum_{a,b = 1}^N
    \int \ud\eta \Big[W(\epsilon_a + \eta, \epsilon_b - \eta| \epsilon_a,
    \epsilon_b)
    \ple_N(\dots, \epsilon_a + \eta, \dots,\epsilon_b - \eta, \dots, t)
    \nonumber\\
    &&  \phantom{\frac{1}{2}\sum_{a,b = 1}^N \int \ud\eta}
    - W(\epsilon_a, \epsilon_b | \epsilon_a - \eta, \epsilon_b + \eta)
    \ple_N(\dots, \epsilon_a, \dots, \epsilon_b, \dots, t)\Big]\,,
    \label{mastereq}
  \end{eqnarray}
\end{widetext}
where the expression of $W$ can be obtained by direct computation of
the collision integrals, which, after rescaling the time variable to
the units of the frequency of binary collision events and thus
absorbing all the parameters into the time scale, yields the universal
function \cite{Gaspard:2008p341}: 
\begin{equation}
  \label{2dkernel}
  W(\epsilon_a, \epsilon_b | \epsilon_a - \eta, \epsilon_b + \eta) =
  \sqrt{\frac{2}{\pi^3}} \times
  \left\{
    \begin{array}{l@{\quad}l}
      \sqrt{\frac{1}{\epsilon_a}}
      K \left(\frac{\epsilon_b + \eta}{\epsilon_a}\right)\\
      \sqrt{\frac{1}{\epsilon_b + \eta}}
      K \left(\frac{\epsilon_a}{\epsilon_b + \eta}\right)\\
      \sqrt{\frac{1}{\epsilon_a - \eta}}
      K \left(\frac{\epsilon_b}{\epsilon_a - \eta}\right)\\
      \sqrt{\frac{1}{\epsilon_b}}
      K \left(\frac{\epsilon_a - \eta}{\epsilon_b}\right)
    \end{array}
  \right.,
\end{equation}
whose defition intervals correspond respectively to 
$-\epsilon_b < \eta < -\mathrm{max}(\epsilon_b - \epsilon_a, 0)$,
$-\mathrm{max}(\epsilon_b - \epsilon_a, 0) < \eta<0$, 
$0 < \eta< \mathrm{max}(\epsilon_a - \epsilon_b, 0)$,
and $\mathrm{max}(\epsilon_a - \epsilon_b, 0) <\eta<\epsilon_a$. Here $K(m)$ 
denotes the complete elliptic intergal of the first kind
\cite{Abramowitz:1970p13721}. 

A system of $N$ isolated cells whose time evolution is specified by
the master equation (\ref{mastereq}) reaches a microcanonical
equilibrium state whose total energy can be parametrized in terms of
the temperature according to $\epsilon_1 + \dots + \epsilon_N = N
T$. The corresponding energy exchange frequency is 
\begin{equation}
  \nu_N(T) = \sqrt{T}\Big[1 + \mathcal{O}(1/N)\Big],
  \label{nut}
\end{equation}	
whose infinite-size limit is simply $\nu(T) = \lim_{N\to\infty}
\nu_N(T) = \sqrt T$ (in the chosen units of time). 

An expression of the heat conductivity of the system described by
Eq.~(\ref{mastereq}) is obtained by considering the Einstein-type
relation satisfied by the variance of the associated Helfand moment,
which measures the spread of energy as a function of time.  

For the system of $N$ energy cells aligned along a one-dimensional
ring, the Helfand moment is defined according to 
\begin{equation}
  H_N(t) = \sum_{i=1}^N i \epsilon_i(t),
  \label{helfand} 
\end{equation}
where $\epsilon_i(t)$ is the state of the energy at site $i$ at time
$t$. This quantity evolves in time by discrete steps, when energy
exchanges occur. Let $\{\tau_n\}_{n\in\mathbb{N}}$ denote the sequence
of times at which successive energy exchanges take place. Assuming
cells $i$ and $i+1$ exchange some amount of energy at time $\tau_n$,
we can write the corresponding change in the Helfand moment as
$\epsilon_i(\tau_{n}-0) - \epsilon_i(\tau_{n}+0)$.  

Computing the mean squared change in the Helfand moment as a function
of time, we obtain an expression of the thermal conductivity according
to 
\begin{equation}
  \kappa(T) = \lim_{N\to\infty} \kappa_N(T),
\end{equation}
where we defined the finite $N$ conductivity to be 
\begin{equation}
  \kappa_N(T) = 
  \frac{1}{N (\kb T)^2}
  \lim_{n\to\infty} \frac{1}{2\tau_n}
  \Big\langle [H_N(\tau_n) - H_N(\tau_0)]^2 \Big\rangle.
  \label{helfandsq}
\end{equation}

In Ref.~[\onlinecite{Gaspard:2008p341}], it was argued that the infinite
$N$ limit of this quantity is determined by the static correlations
only, yielding the result 
\begin{equation}
  \kappa_N(T) = l^2 \sqrt{T}\Big[1 + \mathcal{O}(1/N)\Big].
  \label{kappat}
\end{equation}	

Comparing Eqs.~(\ref{nut}) and (\ref{kappat}), we obtain the announced result
\begin{equation}
  \kappa(T) = l^2 \nu(T),
  \label{kappanumeq}
\end{equation}
which is an exact result for the stochastic system evolved by the
master equation  (\ref{mastereq}) and does not make explicit use of
the form (\ref{2dkernel}), except for some symmetries
\cite{Gaspard:2009p988}. The corresponding result 
(\ref{kappanu}) for the billiard dynamics is obtained by plugging back
the proper timescale of binary collisions and letting $\rhom \to
\rhoc$ so that the separation of timescales (\ref{septimes}) is effective. 

Unlike mass transport in Sinai billiard tables for which the transport
equation (\ref{FPeq}) follows directly from the continuous-time
random walk (\ref{meq}), so that there only remains the problem of
comparing the diffusion coefficient of the billiard to the dimension
formula (\ref{diffcoeff}) in the appropriate parameter regime, the
problem of computing the heat conductivity associated with heat
transport in billiard systems of many confined particles proceeds in
two separate steps.  

Having carried out the reduction of the billiard's pseudo-Liouville
equation to a master equation for a stochastic energy exchange system,
the first step is to establish the cancellation of dynamical
correlations in such a system, which yields the identity
$\lim_{N\to\infty} \kappa_N/\nu_N = l^2$. This is however a delicate
result \cite{Grigo:2011p17620} and requires in-depth knowledge of the
spectral properties of the master equation (\ref{mastereq}). The
approach we took in Ref.~[\onlinecite{Gaspard:2008p341}], though
equivalent, uses different techniques, based on analyzing the first
few orders of the gradient expansion of the kinetic equation to obtain
the expression of the heat current in terms of the local temperature
gradient in a non-equilibrium stationary state, i.~e. Fourier's law.  

The second step of the program is to go back to the billiard dynamics
and investigate the convergence of the heat conductivity to the binary
collision frequency in the limit $\rhom\to\rhoc$.  

For two-dimensional billiard systems such as the ones shown in
Fig.~\ref{fig.2dlatticeb}, the agreement was found to be rather
satisfactory \cite{Gaspard:2008p334}. For the three-dimensional
billiard systems we turn to now, this agreement is even better. 

\section{Three-dimensional billiard for heat
  transport \label{sec.engy3d}} 

\begin{figure}[thbp]
  \centering
  \includegraphics[width = .35\textwidth]{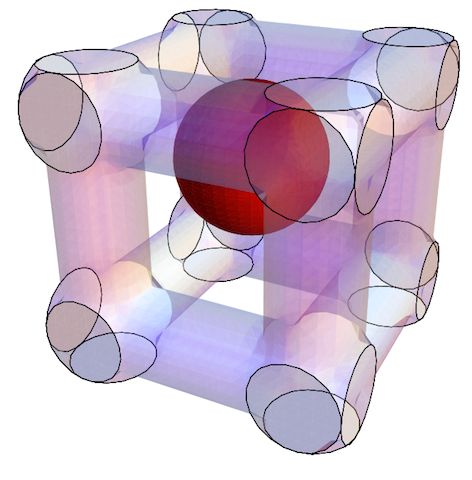}
  \includegraphics[width = .5\textwidth]{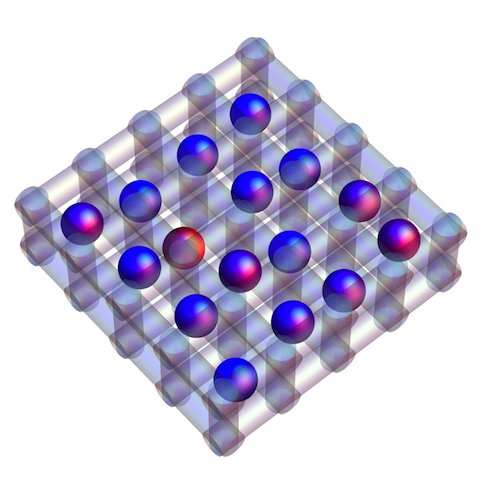}
  \caption{(Top) Hard-sphere particle trapped in a cuboid cell with
    cylindrical edges. (Bottom) A system made out of many copies of
    such cells which form a spatially periodic structure. The cells
    are semi-porous in the sense that particles are prevented from
    escaping, and can yet partially penetrate into the neighboring
    cells, thus allowing energy transfer through collisions among
    neighboring particles. As with the two-dimensional case, the
    likelihood of binary collision events can be controlled by tuning
    the geometry of the cell.} 
  \label{fig.3dlatticeb}
\end{figure}

Consider a three-dimensional billiard of confined hard spheres such as
shown in Fig.~\ref{fig.3dlatticeb}. The reduction of the
pseudo-Liouville equation governing the phase-space evolution of
probability densities $p_N(\{\mathbf{r}_i,\mathbf{v}_i\},t)$ to a
master equation similar to Eq.~(\ref{mastereq}) was already carried
out in Ref.~[\onlinecite{Gaspard:2009p988}]. 

The corresponding kernel, in the appropriate time units, is found to
have the  universal form
\begin{equation}
  W(\epsilon_a, \epsilon_b | \epsilon_a - \eta, \epsilon_b + \eta) =
  \sqrt{\frac{\pi}{8}} \times
  \left\{
    \begin{array}{l}
      \sqrt{\frac{\epsilon_b + \eta}{\epsilon_a \epsilon_b}}\\
      \frac{1}{\sqrt{\mathrm{max}(\epsilon_a, \epsilon_b)}}\\
      \sqrt{\frac{\epsilon_a - \eta}{\epsilon_a \epsilon_b}}
    \end{array}
  \right.,
  \label{3dkernel}
\end{equation}
whose definition intervals correspond respectively to 
$-\epsilon_b < \eta < -\mathrm{max}(\epsilon_b - \epsilon_a,0)$,
$ -\mathrm{max}(\epsilon_b - \epsilon_a, 0) 
< \eta< \mathrm{max}(\epsilon_a - \epsilon_b,0)$, and
$\mathrm{max}(\epsilon_a - \epsilon_b,0)<\eta<\epsilon_a$.

Extensive numerical investigations of the master equation
(\ref{mastereq}) with the stochastic kernel (\ref{3dkernel})
were presented in Ref.~[\onlinecite{Gaspard:2009p988}], supporting
with high precision the validity of Eq.~(\ref{kappanumeq}). Here we
focus on the billiard dynamics and consider the mean squared change in
time of the Helfand moment associated with the distribution of energy
in billiard systems formed by one-dimensional lattices of billiard
cells such as shown in Fig.~\ref{fig.3dlatticeb}.

Let $\{\tau_n\}_{n\in\mathbb{Z}}$ denote the times at successive
binary collision events. As we now need to account for the motion of
particles within their respective cells, there are two types of
contributions to changes in the Helfand moment between two successive
binary collision events. The leading contribution arises from the
energy exchanges 
which take place at binary collision events. Thus, when a binary
collision occurs between particles $j$ and $k$, the Helfand moment
changes by the amount $[x_j(\tau_n) - x_k(\tau_n)]
[\epsilon_j(\tau_{n}+0) - \epsilon_j(\tau_{n}-0)]$, where
$x_{j}(\tau_n)$ and $x_{k}(\tau_n)$ denote the positions of particles
$j$ and $k$ along the direction of the lattice spatial extension at
time $\tau_n$. The other contribution to the Helfand moment arises
from the advection of particles within their respective cells
according to $\sum_a [x_a(\tau_{n}) - x_a(\tau_{n-1})] \epsilon_a
(\tau_{n-1})$.

The range of allowed parameter values is identical to the
two-dimensional case: $l/2 \leq \rho < l/\sqrt 2$, and 
$\rhoc < \rhom < \rho$, where $\rhoc \equiv \sqrt{\rho^2 - l^2/4}$.

Taking $l=1$, we fix $\rho = 0.50$ so $\rhoc = 0$ and use the six
parameter values $\rhom = 0.20, 0.25, \dots, 0.45$. For each one of
them, we fix the total energy to be $E = 3N/2$ ($T=1$) and vary the
system size from $N=3$ and up to 
$N=100$ cells. We typically run $10^3$ trajectories, each for a
duration of $10^3\times N$ units of the average time between collision
events. We compute: (i) the Helfand moment versus time and take the
mean and standard deviation of the transposition of
Eq.~(\ref{helfandsq}) to obtain the conductivities $\kappab_N$; (ii) the
binary and wall collision frequencies $\nub_N$ and $\nuw_N$ by
directly averaging their numbers with respect to time. We then use
linear fits in $1/N$ for $N\geq 10$ of $\kappab_N$ and $\nub_N$ to
obtain the extrapolation $\nub_\infty = \lim_{N\to\infty} \nub_N$ and
$\kappab_\infty = \lim_{N\to\infty} \kappab_N$. Here and below,
explicit temperature dependencies are dropped.

\begin{figure}[htbp]
  \centering
  \includegraphics[width = .45\textwidth]{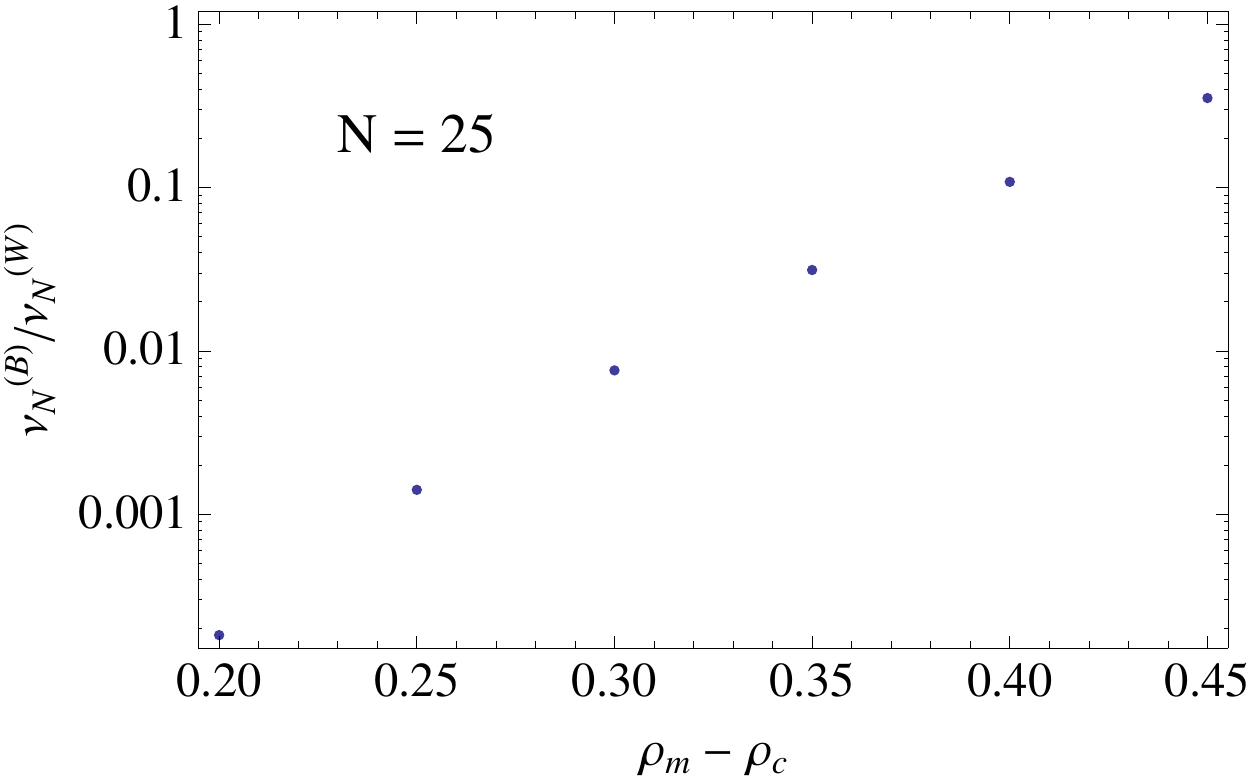}
  \caption{Ratio between the binary and wall collision frequencies as
    a function of the parameter $\rhom$ for $N=25$ ($\rho = 0.50$,
    $\rhoc = 0$).} 
  \label{fig.nuBnuW}
\end{figure}

\begin{figure}[htbp]
  \centering
  \includegraphics[width = .45\textwidth]{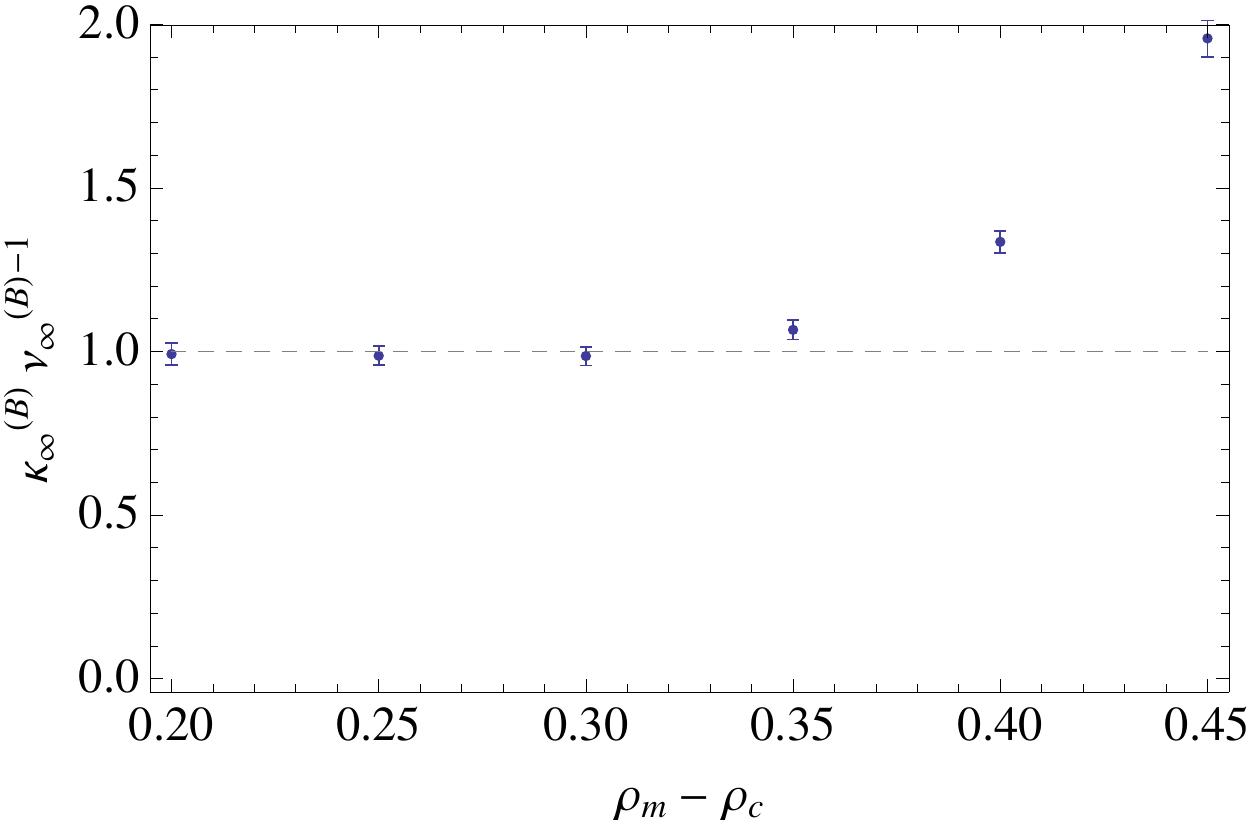}
  \caption{Ratio between the heat conductivity of the infinite length
    system and the corresponding collision frequency between neighboring
    particles as a function of the parameter $\rhom$ ($\rho = 0.50$ and
    $\rhoc = 0$). The error bars show the widths of the computed
    0.95 confidence intervals.} 
  \label{fig.kappanuvrhom}
\end{figure}

\begin{figure*}[hptb]
  \centering
  \includegraphics[width = .45\textwidth]{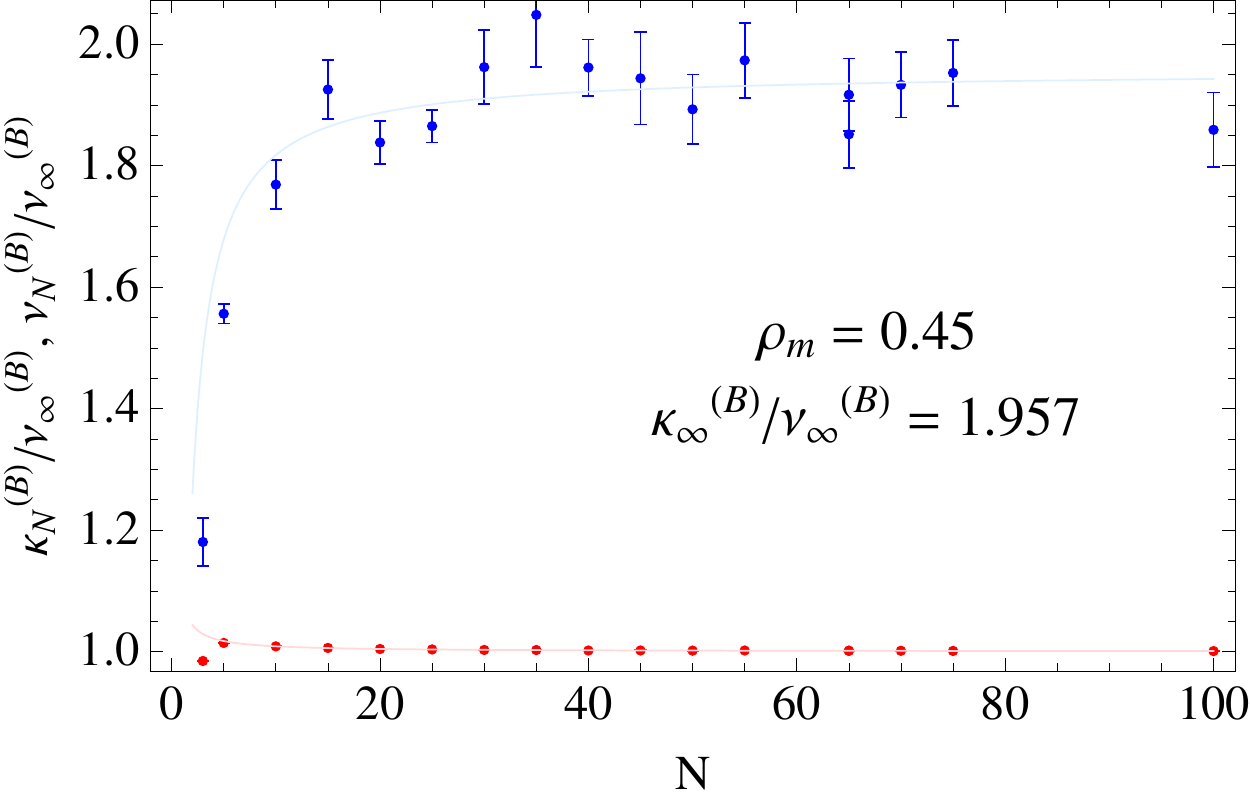}
  \includegraphics[width = .45\textwidth]{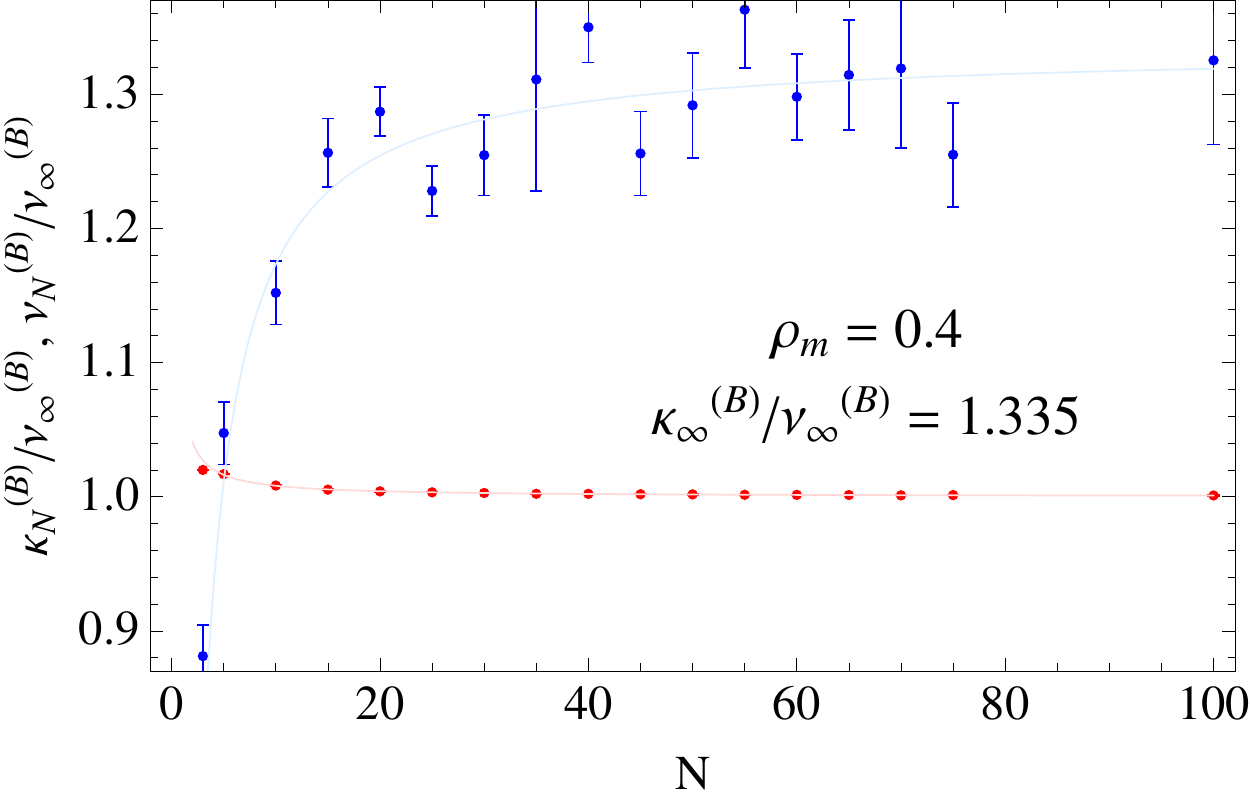}
  \includegraphics[width = .45\textwidth]{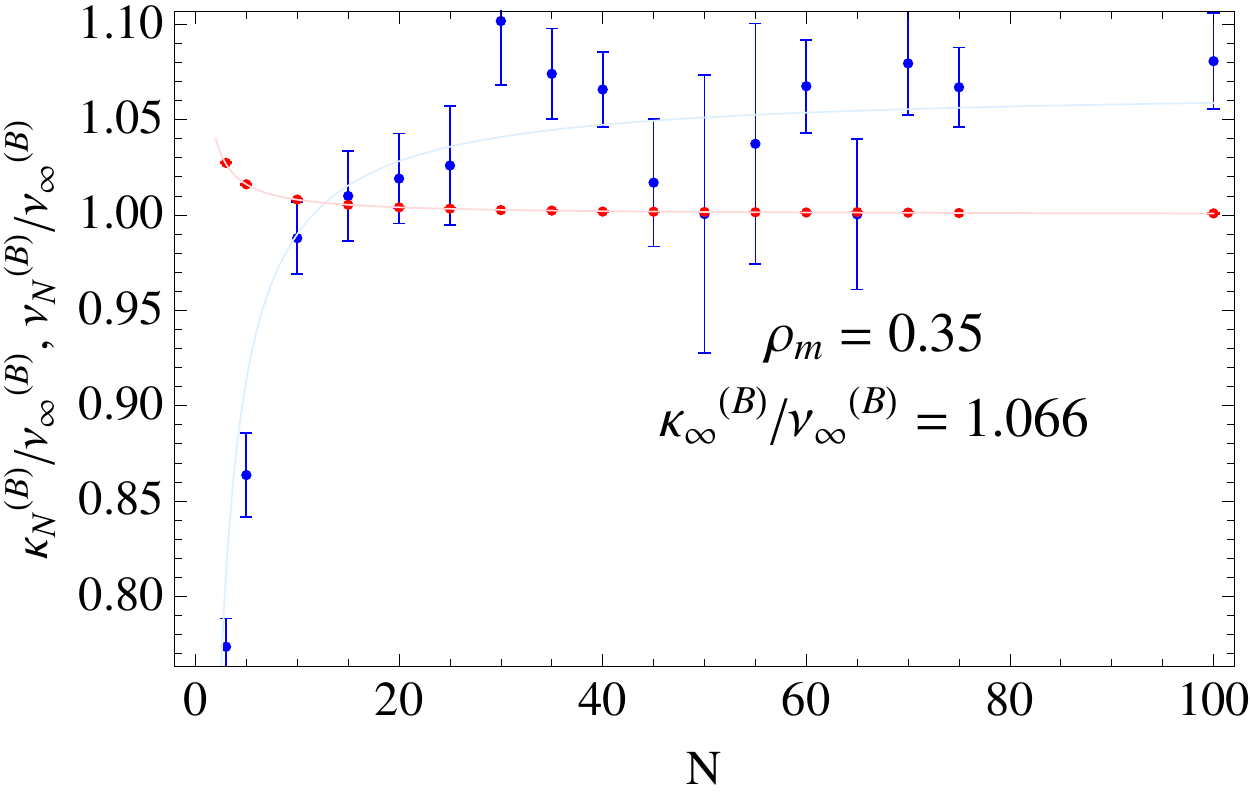}
  \includegraphics[width = .45\textwidth]{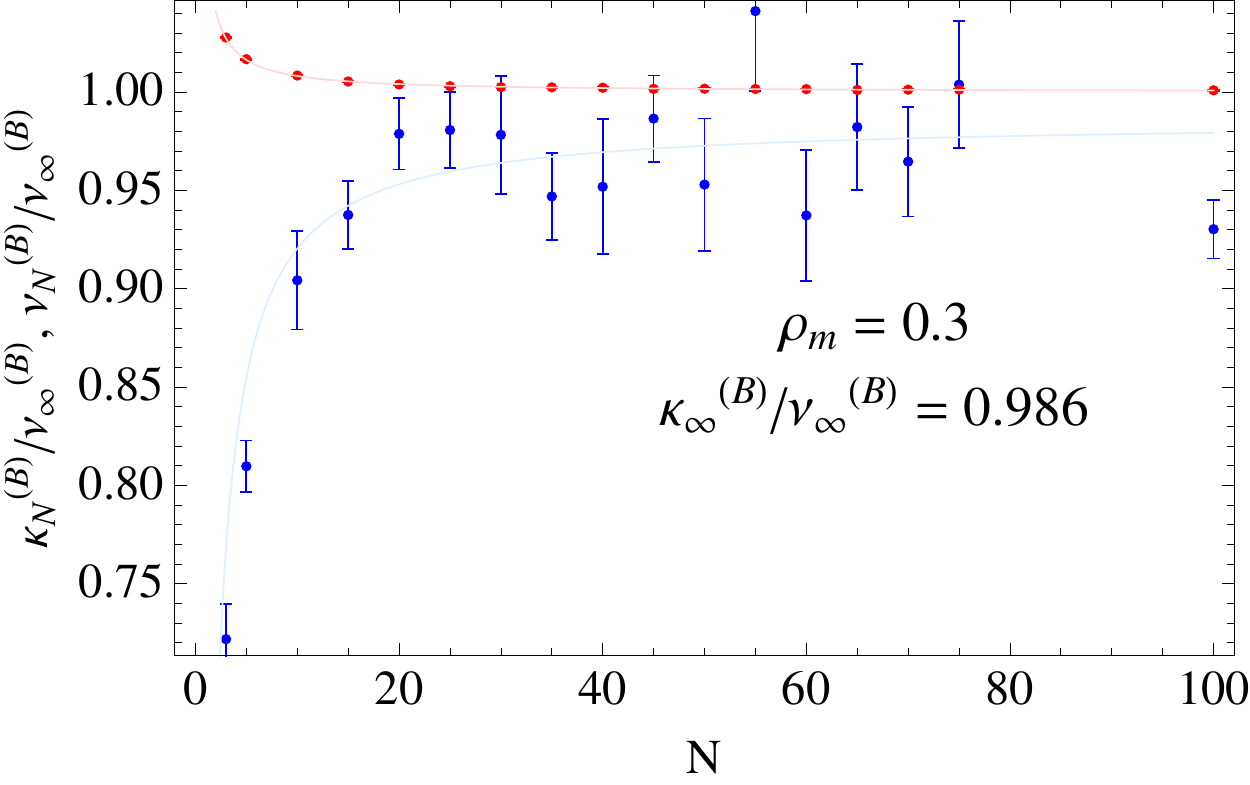}
  \includegraphics[width = .45\textwidth]{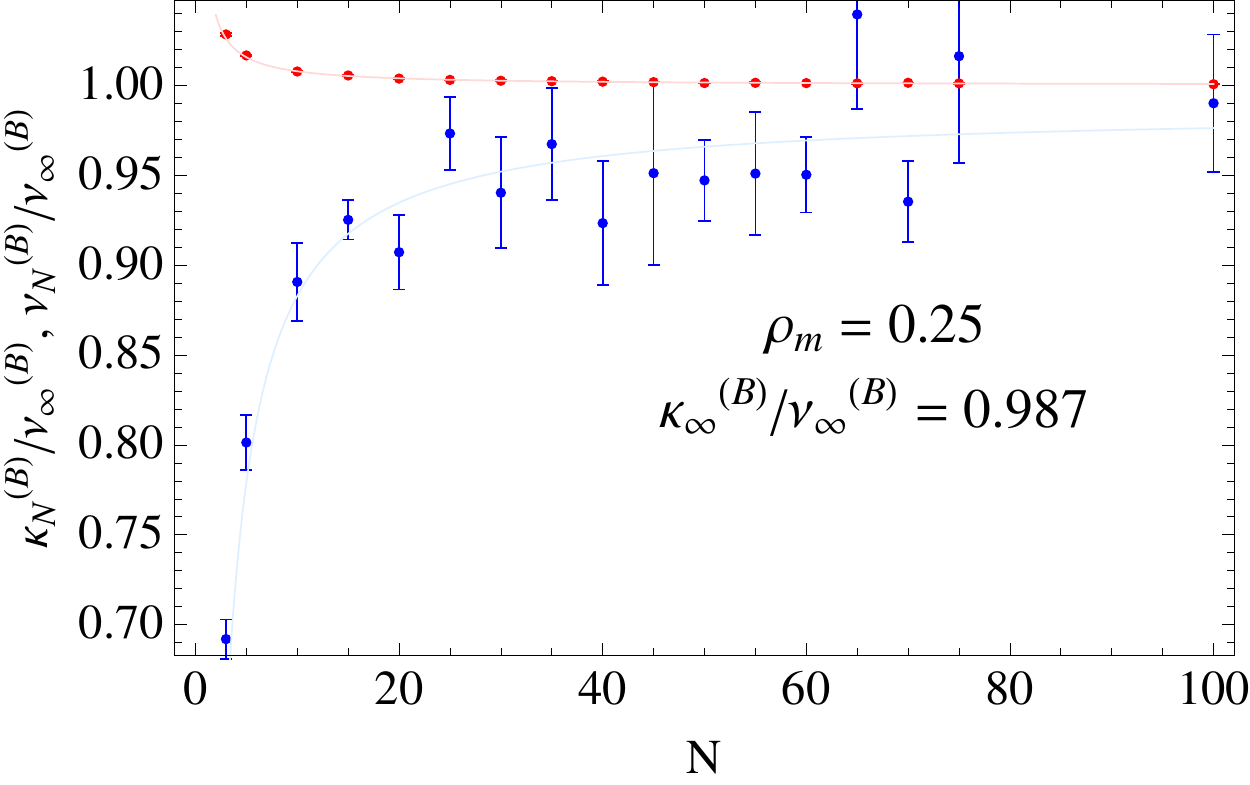}
  \includegraphics[width = .45\textwidth]{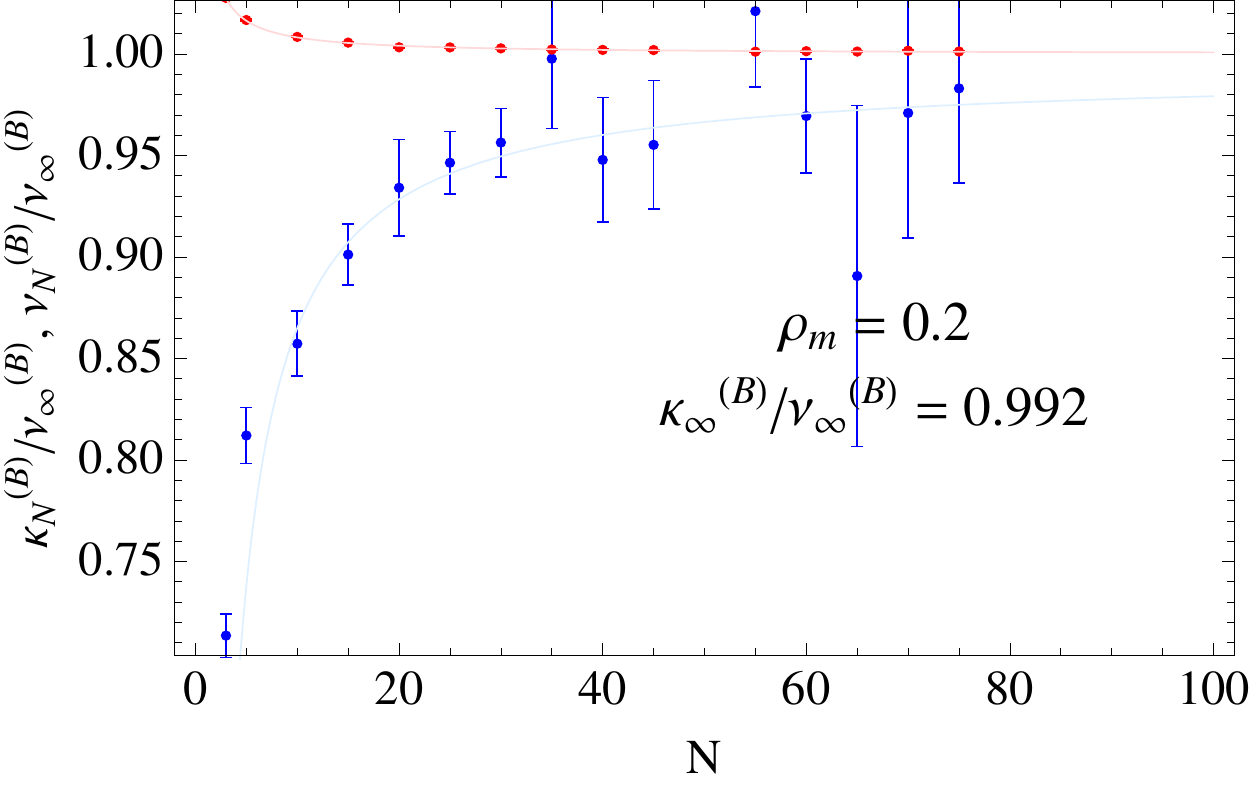}
  \caption{Heat conductivities $\kappab_N$ and binary collision frequencies
    $\nub_N$ measured as functions of the system size $N$ for
    different values of the parameter $\rhom$ ($\rho = 0.50$). The
    solid curves correspond to the linear fits of $\kappab_N$ and
    $\nub_N$ as functions of $1/N$. The intercepts yield the infinite
    size estimates $\kappab_\infty$ and $\nub_\infty$. } 
  \label{fig.kappanuvN}
\end{figure*}

Measurements of the ratio between $\nub_N$ and $\nuw_N$, by which we
can assess the effectiveness of the separation of timescales
(\ref{septimes}), are displayed in Fig.~\ref{fig.nuBnuW} for $N=25$
(other system sizes yield similar values). The results of the
computations of $\kappab_\infty/\nub_\infty$ for the different
parameter values $\rhom$ are displayed in Fig.~\ref{fig.kappanuvrhom}
and are found to be in very good agreement (within two digits) of the
dimensional formula (\ref{kappanu}) for values of $\rhom$ as large as
$0.30$. In Fig.~\ref{fig.kappanuvN}, the details of the fitting
procedure used to obtain the values of $\kappab_\infty$ and
$\nub_\infty$ are shown for the different parameter values as
functions of $N$.

\section{Concluding Remarks\label {sec.conc}}

By controlling the rate at which a tracer hops from cell to cell in a
Sinai billiard table or that of interaction among neighboring disks or
spheres in a high-dimensional billiard with local confinement rules,
one identifies a limit of vanishing rate where the complicated
phase-space dynamics are replaced by stochastic processes which
account for the transport of mass in the first case, or (kinetic)
energy in the second. Furthermore, the transport coefficients of these
stochastic processes have the same simple dimensional expression,
given by the length scale of transfers squared multiplied by the
corresponding rates. 

As emphasized in this paper, the dimensional formulae we obtained for
the transport coefficients of the  models of mass and heat transport we
have considered are but two faces of the same coin. Indeed, in both
cases the accuracy of our approximation of the transport coefficients
of the billiards in terms of hopping or collision rates relies on the
efficient separation of two timescales, namely the timescale of local
collision events must be much shorter than that characterizing the
transfer of mass or energy. In other words, the relaxation to local
equilibrium precedes mass or energy transfers. 

We can in fact view the notion of relaxation to local equilibrium as a
low-dimensional transposition of that of local thermal equilibrium,
which is at the heart of many theories involving hydrodynamic scaling
limits and typically assumes a large number of degrees of freedom
\cite{Bonetto:2000p13477}. In our billiards, the relaxation to local
equilibrium occurs on the constant energy surface of a single
particle. But one could instead consider systems of trapped gases, in
which case the relaxation to local equilibrium would involve transfers
of energy among the particles in the same trap. In such a case,
provided relaxation to local equilibrium takes place on timescales
much shorter than that of energy transfer between neighboring traps,
a similar dimensional expression of the heat conductivity of the
corresponding stochastic process in terms of the product of length
scale squared and rate of energy transfer would yield an accurate
approximation of the transport coefficient of the billiard system. 

We end with a remark concerning the closure (\ref{closure}) which
relies on the ergodicity of the local dynamics on the surface of
constant energy. As noted already by Zwanzig \cite{Zwanzig:1983p33},
Sinai billiards cannot be replaced by polygonal ones, for which the
notion of ergodicity is weaker since the velocity directions take on
values in a discrete set. In energy exchange processes, however,
ergodicity of the many-particle billiard may be restored through the
sole interaction among neighboring particles. As shown in
Ref.~[\onlinecite{Gilbert:2008p354}], an elastic string-type interaction
between particles trapped in polygonal boxes provides a simple model
of a system which, on the one hand, cannot be accurately described by a
master equation similar to Eq.~(\ref{mastereq}), even when
interactions are rare, but on the other hand, has a well-defined heat
conductivity which is well approximated by a dimensional formula. An
understanding of the transport properties of this system beyond the
Boltzmann hypothesis remains to be elucidated.

\begin{acknowledgments}
We dedicate this paper to the memory of Sasha Loskutov in appreciation
for his co-organizing the conference Billiards'2011 in Ubatuba, SP,
Brazil. TG also wishes to thank E. Leonel for his warm hospitality.
The authors would further like to thank F. Barra, L. Bunimovich,
R. Lefevere, M. Lenci, C. Liverani, V. Rom-Kedar, D. P. Sanders and
D. Sz\'asz for stimulating discussions which took place at different
stages of this project. They acknowledge financial support by the 
Belgian Federal Government under the Interuniversity Attraction Pole
project NOSY P06/02 and FRS-FNRS under contract C-Net NR/FVH 972. TG
is financially supported by the Fonds de la Recherche Scientifique
FRS-FNRS and receives additional support through FRFC convention
2,4592.11.
\end{acknowledgments}

%

\end{document}